\documentstyle[preprint,aps]{revtex}
\begin{document}

\title{Influence of a magnetic field on the antiferromagnetic order in
UPt$_3$}

\author{B. Lussier$^1$, L. Taillefer$^{1,*}$,
W.J.L. Buyers$^{2,*}$, T.E. Mason$^{3,*}$
and T. Petersen$^3$}

\address{$^1$Department of Physics, McGill University, Montr\'{e}al, Qu\'{e}bec, Canada, 
H3A 2T8\\
$^2$AECL, Chalk River, Ontario, Canada, K0J 1J0\\
$^3$Department of Physics, University of Toronto, Toronto, Ontario, Canada, M5S 1A7\\
$^*$Canadian Institute for Advanced Research}
 
\date{\today}

\maketitle

\begin{abstract}

A neutron diffraction 
experiment was performed to investigate the effect of a magnetic field
on the antiferromagnetic order in the heavy fermion superconductor UPt$_3$.
Our results show that a 
field in the basal plane of up to 3.2 Tesla, higher than H$_{c2}$(0), 
has no effect: 
it can neither select a domain nor rotate the moment.
This has a direct impact on current theories for the superconducting phase diagram based on a
coupling to the magnetic order.
 
\end{abstract}
 

\tighten
Most of the heavy fermion superconductors order antiferromagnetically
before the onset of superconductivity, with T$_N$ $\simeq$  10 T$_c$.
The possible relation between the phenomena is one of the central issues in the field.
However, no two compounds have exactly the same magnetic behavior. While both 
UPt$_3$ \cite{Aeppli et al. 1988} and URu$_2$Si$_2$ \cite{URu2Si2
neutrons} show an extremely
small ordered moment, of order 0.01 $\mu_B$/U atom, it is as large as 0.85 $\mu_B$/U atom in
UPd$_2$Al$_3$ \cite{UPd2Al3 moment}. The specific heat anomaly at
T$_N$ is large in URu$_2$Si$_2$ \cite{Palstra et al. 1986} yet 
absent in UPt$_3$ \cite{Fisher et al. 1991}. The ordered structure breaks the hexagonal symmetry in UPt$_3$ and
UPd$_2$Al$_3$, with the moments aligned in the basal plane, while the tetragonal symmetry of
URu$_2$Si$_2$ is preserved. The magnetic order and fluctuations are unaffected 
by the onset of superconductivity
in UPd$_2$Al$_3$ \cite{UPd2Al3 moment and SC}, while a slight
decrease in the amplitude of the moment is observed in UPt$_3$
\cite{Aeppli et al. 1989,Isaacs et al. 1995} and a
saturation of the moment in URu$_2$Si$_2$ \cite{Mason et al. 1990}.
 
The coexistence of magnetism and superconductivity in these compounds
has been viewed as evidence for an unconventional pairing mechanism.
Unlike the Chevrel phases, where the electrons
responsible for the superconductivity are distinct from 
those responsible for the magnetism, 
it appears that in the case of UPt$_3$, in particular,
the same electrons participate in both phenomena.
Indeed, in this material a division of labor is implausible in view of the
presence of the f-electrons at the Fermi level and the fairly uniform 
effective mass around the Fermi surface \cite{Taillefer and Lonzarich
1988,Norman et al. 1988}.
 
The unconventional nature of the superconducting state in UPt$_3$ is
most strikingly manifest in the existence of several
superconducting phases \cite{Taillefer 1994}. The magnetic field (H)-pressure (P)-
temperature (T) phase diagram shows two distinct transitions at
T$_c^+$=0.5 K and T$_c^-$=0.44 K for H=P=0 \cite{Fisher et al. 1989}. Application of a magnetic field in the basal
plane ($\vec{H} \perp \hat{c}$) brings the two transitions together at a tetracritical point
\cite{Hasselbach et al. 1989}, which shows up clearly on the
H$_{c2}$(T) line as a kink at a field H$^*$ of about 0.4 Tesla \cite{Kink}.
Hydrostatic pressure also causes T$_c^+$ and T$_c^-$ to merge, at a critical pressure of about
3.7 kbar \cite{Trappmann et al. 1991}. A complete theory for the phase diagram of UPt$_3$ has been one of the major
pursuits in the field over the past five years. Two main 
scenarios are currently debated:
in the first type, the proximity of T$_c^+$ and T$_c^-$ is considered
accidental and the 
two zero-field phases are attributed to different representations 
of the order parameter \cite{Garg 1993}.  In the second type, the double transition is viewed 
as a splitting resulting from the lifting of the degeneracy of a
state (within a single
representation for the order parameter) by 
some symmetry-breaking field \cite{Machida Joynt,Sauls 1994}.  An obvious choice for such a field is the
antiferromagnetic order, with its moment and propagation vector both lying in the basal
plane ($\vec{M_s}$ $\|$ $\vec{q}$ $\|$ $\hat{a}^*$).
The moment configuration
has been described so far in terms of a single-$\vec{q}$ structure
with a given sample in general possessing three
equivalent domains \cite{Aeppli et al. 1989,Isaacs et al. 1995,Hayden
et al. 1992}.  However, the existing data is also compatible with
a triple-$\vec{q}$ structure.
 
In their neutron study under 
pressure, Hayden {\it et al.} \cite{Hayden et al. 1992} found 
that the antiferromagnetic moment of UPt$_3$ is fully suppressed by
applying 3 to 4 kbar, which is also the critical pressure for
the merging of T$_c^+$ 
and T$_c^-$. The parallel disappearance of magnetism and phase 
multiplicity under pressure is strong
evidence in favor of the coupling scenarios 
(the second type), with the antiferromagnetic order acting as the 
symmetry-breaking field.
Within the coupling scenarios, the kink in the H$_{c2}$ curve is basically the result of a
sudden reorientation of the (vector) order parameter $\vec{\eta}$ in the basal plane \cite{Machida Joynt}. 
Both the moment $\vec{M_s}$ and
the field $\vec{H}$ will couple to $\vec{\eta}$, each trying to align it in the minimum
energy direction. Without loss of generality, let us consider
the case of $\vec{M_s} \perp \vec{H}$, with both couplings to
$\vec{\eta}$ favoring {\it parallel} alignment. At low fields, the coupling to the magnetic
order dominates and
$\vec{M_s}$ determines the orientation of $\vec{\eta}$. Then, 
when the field is increased to the point where its coupling dominates,
a reorientation of $\vec{\eta}$ occurs, causing a kink in H$_{c2}$(T). Of course, if the
field direction is instead made parallel to $\vec{M_s}$, no kink is predicted, 
since there is no competition between the two couplings. As a result,
within a single antiferromagnetic
domain (assuming a single-$\vec{q}$ structure for the magnetic order)
the upper critical field in the basal plane
of UPt$_3$ is predicted to show
a sharp kink only for one direction of the field (say $\vec{H}$ $\|$
 $\hat{a}$), and no kink for
the $\hat{a}^*$ direction 90$^\circ$ away \cite{Machida Joynt}.
Experimentally, however,
a kink is observed at H$^* \approx$ 0.4 Tesla 
for any high-symmetry
direction (0$^\circ$, 90$^\circ$, 120$^\circ$ relative to $\hat{a}$) 
\cite{Ubiquitous}. The theory can be reconciled with a ubiquitous kink by
supposing that the moment is not fixed to the lattice but rather follows 
the field in
such a way that $\vec{M_s} \perp \vec{H}$ for all field orientations in 
the basal plane.
This is possible provided
the in-plane magnetic anisotropy energy is negligible 
compared to the Zeeman energy acting on $\vec{M_s}$.
Sauls \cite{Sauls 1995} 
showed that a rotation of $\vec{M_s}$ in the basal plane is accompanied
by a modulation of its amplitude $M_s$ with 60$^\circ$ periodicity, which in turn causes
H$_{c2}$($\theta$) to exhibit 60$^\circ$ oscillations, such as those observed recently in
UPt$_3$ \cite{Keller et al. 1994}. 
The first goal of our experiment was to determine
whether a magnetic field lower than one Tesla can indeed cause the magnetic
moment to rotate in the basal plane away from its zero-field
configuration ($\vec{M_s}$ $\|$ $\vec{q}$ $\|$ $\hat{a}^*$) and
remain perpendicular to $\vec{H}$.

If the magnetic ground state of UPt$_3$ has only one propagating vector (single-$\vec{q}$), as assumed
until now by all authors \cite{Aeppli et al. 1988,Aeppli et al.
1989,Isaacs et al. 1995,Hayden et al. 1992}, then there
should in general be 3 independent domains
with $\vec{M_s}$ oriented at 120$^\circ$ with respect to each other.
Agterberg and Walker \cite{Agterberg and Walker 1995a} 
have recently considered the effect of having 3 possible domains on the H$_{c2}$ curve of
UPt$_3$ in the basal plane. They assume that $\vec{M_s}$ is fixed with respect to the crystal
lattice (i.e. parallel to any one of the 3 a*-axes) but that only the most thermodynamically stable domain will be
populated for any given field direction. Within the coupling scenario, the implications are
fairly straightforward: the angle between $\vec{M_s}$ and $\vec{H}$ can only range over
$\pm$30$^\circ$ and the domain selection by the field as it is rotated causes a 60$^\circ$
variation in H$_{c2}$(T). The limited range of angles could perhaps
explain why a straight H$_{c2}$ curve is never observed.
The second goal of our experiment was therefore
to establish whether a magnetic field of less than one Tesla
can select a single domain.

We show that a magnetic field of up to 3.2 Tesla in the 
basal plane -- 
which is greater than H$_{c2}$(0) and much greater than H$^*$ -- has no influence on
the antiferromagnetic order: it can neither rotate the moments nor
select a domain.
 

Our neutron diffraction studies were performed with the DUALSPEC
triple-axis spectrometer at the NRU reactor at Chalk River Laboratories with a 
pyrolytic graphite 
monochromator, analyzer and filter, and a neutron wavelength of 2.37 \AA.  
The collimation was 0.6$^\circ$ between the 
monochromator and sample and 0.8$^\circ$ between sample and analyzer.  The 
sample, used in previous neutron experiments \cite{Hayden et al. 1992}, 
was a high-quality single 
crystal of UPt$_3$ that exhibits two sharp successive 
superconducting transitions, a moment of 0.03 $\mu_B$/U atom and a N\'eel 
temperature of
approximately 6 K.  It was 
aligned with its hexagonal plane in the scattering plane of
the spectrometer and mounted in a horizontal field cryostat that 
enabled a field of up to 3.2 Tesla to be applied at any angle in the 
basal plane.
 
 
In a first measurement, the magnetic field was applied in the
basal plane along the [$\bar{1}$, 2, 0] direction, which is
perpendicular to the a$^*$ direction and to the wave vector
of the $\vec{q_1}$ = ($\frac{\rm 1}{\rm 2}$, 0, 0) domain.  This
should favor the $\vec{q_1}$ domain and remove the $\vec{q_2}$
= ($\bar{\frac{\rm 1}{\rm 2}}$, $\frac{\rm 1}{\rm 2}$, 0) and 
$\vec{q_3}$ = (0,$\bar{\frac{\rm 1}{\rm 2}}$, 0) domains, each of
which is at 30$^\circ$  to the applied field.  The intensity of the
$\vec{q_1}$ peak, observable at a scattering wave vector 
$\vec{Q_1}$ =($\frac{\rm 1}{\rm 2}$, 1, 0), which is at an
angle to $\vec{M_1} \| \vec{q_1}$ in order to sense the
moment (see Fig. 1), should then increase by a factor three on application of
a sufficiently strong field.  Concomitantly, the intensities of the 
$\vec{q_2}$ domain at $\vec{Q_2}$=($\bar{\frac{\rm 3}{\rm 2}}$,
$\frac{\rm1}{\rm 2}$, 0) and the $\vec{q_3}$ domain at 
$\vec{Q_3}$ = ($\bar{1}$, $\frac{\rm 3}{\rm 2}$ , 0) should vanish.
 
From scans such as those
displayed in Fig. 2, in which the crystal angle $\psi$ was rotated 
through the Bragg position at a fixed temperature of 1.8 K and a fixed 
field orientation, namely
$\vec{H} \perp \vec{q_1}$,
we find that the Bragg peaks 
corresponding to the three wave vectors persist up to a magnetic 
field of 3.2 Tesla, as shown in Fig. 3.
There is
no significant increase in the population
of what should be the most thermodynamically stable
domain ($\vec{q_1}$).  A slight increase of order 30\% at 3 Tesla
is not inconsistent with the error bars in Fig. 3.  This would then
be compatible with a roughly equivalent decrease observed in the $\vec{q_2}$
intensity, and suggest that complete domain repopulation could be achieved at
higher fields.  However, as far as the superconducting
phase diagram is concerned, it is important to stress that this anisotropy field
is larger than $H_{c2}(0)$, so that the sample is multi-domain in
all superconducting phases.

In order to make $\vec{q_2}$ the least favored domain,
we rotated the field by 30$^{\circ}$ to lie along the $\vec{q_2}$ 
direction.
At 1.6 Tesla, we again observed that both the $\vec{q_1}$ and $\vec{q_2}$ 
modulations remain present.
Within the statistical error of 20\%,
the integrated intensity of the $\vec{q_2}$ modulation observed at a 
scattering vector $\vec{Q_2}$=($\bar{\frac{\rm 3}{\rm 2}}$,$\frac{\rm
1}{\rm 2}$,0) was unchanged between 0 
and 1.6 Tesla.  For independent (and weakly pinned) domains the intensity would 
have vanished. A similar independence of field was observed for the 
$\vec{q_1}$ modulation seen at $\vec{Q_1}$=($\frac{\rm 1}{\rm
2}$,1,0), where the peak 
should have grown by a factor of $\frac{\rm 3}{\rm 2}$.

This is in contrast with the behavior of UPd$_2$Al$_3$ \cite{Mason
1995}, where a
field of less than one Tesla in the hexagonal
basal plane perpendicular to $\vec{q}$=(1,1,0) 
clearly enhances the population of that particular domain 
to the detriment of the other two. If a similar effect occured in
UPt$_3$, the relative intensities of the $\vec{q_1}$ and $\vec{q_2}$
domains would be expected to follow the solid lines shown in Fig.
3.
 
In UNi$_2$Al$_3$, where the moment is 0.12 $\mu_B$/U atom,
intermediate between that of
UPt$_3$ and that of UPd$_2$Al$_3$, the propagation vector 
(0.61,0,0.5) also has a component in the basal plane but it
is incommensurate with the crystal lattice \cite{Schroder et al.}. 
In this case, a field of 3 Tesla is insufficient
to produce a monodomain \cite{Gaulin 1995}.

In zero field cooled (ZFC) experiments, such as those described above, it is
possible that domains, having already formed, cannot attain the new
thermodynamic equilibrium associated with the applied field.  To check
for this possibility, we slowly cooled
the sample through its 6 K magnetic transition 
in a field of 3.2 Tesla
along the ($\bar{1}$,2,0) direction. 
All three wave vector 
modulations were found to have condensed with the same intensity as for 
cooling in zero field.  For the ${\vec q_1}$
modulation we can exclude at 
the 2$\sigma$ level any increase in peak intensity beyond 30\%
relative to the ZFC intensities; field 
selection of one domain would have produced a three-fold intensity 
increase.  These results exclude the possibility that an 
energy barrier, arising from the reduced
orthorhombic symmetry of single-$\vec{q}$
ordered state, might have prevented the 
attainment of an equilibrium domain configuration at low temperature.  
We therefore conclude that in UPt$_3$ the three modulations are present
with roughly equal importance for all
field strengths at which the superconducting state exists.
 
Even if all three wave vectors survive the application of a magnetic 
field, the moments themselves might still
rotate away from being longitudinal ($\vec{M_s}$ $\|$ $\vec{q}$). 
To test this possibility, we monitored the scattering wave vector
$\vec{Q}$ =($\bar{\frac{\rm 1}{\rm 2}}$,$\frac{\rm 1}{\rm 2}$,0),
where neutron
diffraction senses the $\vec{q_2}$ spatial periodicity, but 
where, in the absence of a field, the scattering amplitude is zero 
because the moment is parallel to $\vec{Q}$.  Moment canting 
in the field would then give a non-zero amplitude.  Applying a 
field of 2.8 Tesla along ($\bar{1}$,2,0) -- perpendicular to ${\vec 
q_1}$ and at 30$^\circ$ to $\vec{q_2}$ -- we 
observed no measurable growth in intensity above background.  The
statistics allow us to put
an upper-bound
of 26$^\circ$ on any rotation at the $\sigma$ confidence level (a
realignment of the $\vec{M_s}$ moment of domain $\vec{q_2}$ by the
field would have meant a 60$^\circ$ rotation).
This shows that the moment does not follow the field as 
the latter is rotated in the basal plane, and this for field
strengths much greater
than H*=0.4 Tesla.
This suggests that $\vec{M_s}$ is strongly 
coupled to the crystal lattice, in agreement with the observation
that $\vec{M_s}$ does not rotate upon entering
the superconducting state at 0.5 K
\cite{Isaacs et al. 1995}.

Let us look more closely at the single-$\vec{q}$ assumption.
Isaacs {\it et al.}
\cite{Isaacs et al. 1995} have shown that a collinear
structure with three separate domains gives a diffraction pattern 
consistent with the observed structure factors. 
The question is:
why are all 3 domains equally favored upon cooling in a 
field of 3.2 Tesla which is only perpendicular to one of the associated 
moments?  For a collinear antiferromagnet, the fact that the transverse
susceptibility is larger than the longitudinal susceptibility should lead
to the selection of the domain perpendicular to the applied magnetic field, as
is seen in UPd$_2$Al$_3$. 
A simple explanation for the ubiquitous presence of all 
3 wavevectors is that
the magnetic structure might be 
triple-$\vec{q}$. With a symmetric superposition of three 
equivalent modulations, the diffraction pattern would be the same as
with three single-$\vec{q}$ domains.
A magnetic field would  
have no effect at low fields; it would only produce a single-$\vec{q}$ 
domain sample when the Zeeman energy developing from distortion of the 
3-$\vec{q}$ structure exceeded the binding energy of the 3-$\vec{q}$ 
state.
Triple-$\vec{q}$ structures are known to occur in uranium compounds, such
as USb \cite{Rossat-Mignod et al.} and UPd$_3$ \cite{UPd3},
and are characterized by an insensitivity to
applied magnetic fields and uniaxial stress \cite{Rossat-Mignod et al.}.
Now, it is far from obvious that such a magnetic order could
break the hexagonal symmetry (in zero field), and even more so that
a coupling to the superconducting order can lead to a split transition.
Therefore, if such a structure is the correct one for UPt$_3$, a major reassessment of
the coupling theories mentioned above is needed.

In conclusion, we have shown that basal plane magnetic fields of up to 
3.2 Tesla have no effect on the
magnetic order in UPt$_3$, whether it be in rotating the moments or
in selecting a domain with a single wave vector.
Because the upper critical field of UPt$_3$ is less than 3.2 Tesla,
the absence of rotation
makes it difficult to reconcile the fact that experimentally a kink in H$_{c2}$(T) is
observed at 0.4 Tesla \cite{Hasselbach et al. 1989,Kink,Ubiquitous,Keller et
al. 1994} for various field directions
in the basal plane with the prediction of current
theories \cite{Machida Joynt,Sauls 1994,Sauls 1995}
that it should only occur for one direction of $\vec{H}$ with respect
to $\vec{M_s}$.
In this respect, a calculation with {\it three} fixed domains would prove helpful.
Our results also
invalidate the respective assumptions (moment rotation and domain
selection) underlying two recent explanations
\cite{Sauls 1995,Agterberg and Walker 1995b} for the slight 60$^\circ$
variation of H$_{c2}$ in the basal plane \cite{Keller et al. 1994}.
Finally, there is a distinct possibility that the antiferromagnetic order
in UPt$_3$ has a triple-$\vec{q}$ structure, as opposed to the
single-$\vec{q}$ structure assumed until now, which would require
a major reassessment of
current theories for the superconducting phase diagram.

This work was funded by NSERC of Canada, FCAR of Qu\'ebec and 
the Canadian Institute for Advanced Research. L.T. 
acknowledges the support of the A.P. Sloan Foundation.

\begin{figure}
\caption{Reciprocal space diffraction geometry for the two domains
investigated here.
The {\bf q$_i$}
and {\bf Q$_i$} 
indicate the propagation and
scattering vectors, respectively.
}
\label{fig1}
\end{figure}

\begin{figure}
\caption{Magnetic Bragg peaks at $\vec{q_1}$ and $\vec{q_2}$ 
for H = 0 and 2.8 Tesla, with $\vec{H} \perp \vec{q_1}$.  Complete
selection of a
single domain by the 2.8 Tesla field would eliminate the $\vec{q_2}$ 
Bragg peak and increase the intensity of the $\vec{q_1}$ peak by
a factor 3.
}
\label{fig2}
\end{figure}
 
\begin{figure}
\caption{Integrated intensity as a function of field for 
$\vec{q_1}$ (open circles) and $\vec{q_2}$ (solid circles) with $\vec{H} 
\perp \vec{q_1}$.   The solid lines show the expected behavior for
both magnetic domains for an anisotropy field of order 0.5 Tesla (as
observed in UPd$_2$Al$_3$ [25]).
}
\label{fig3}
\end{figure}

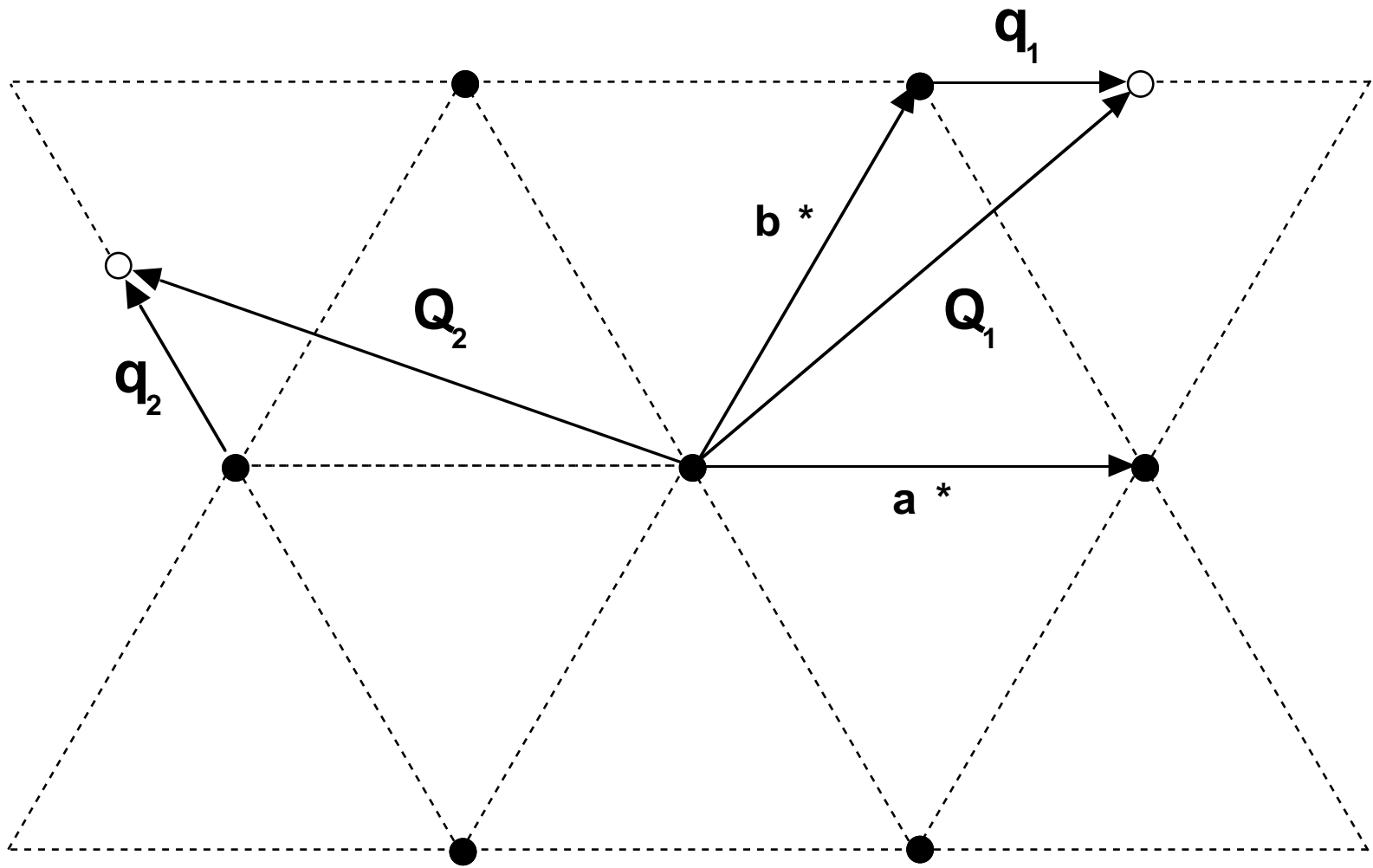
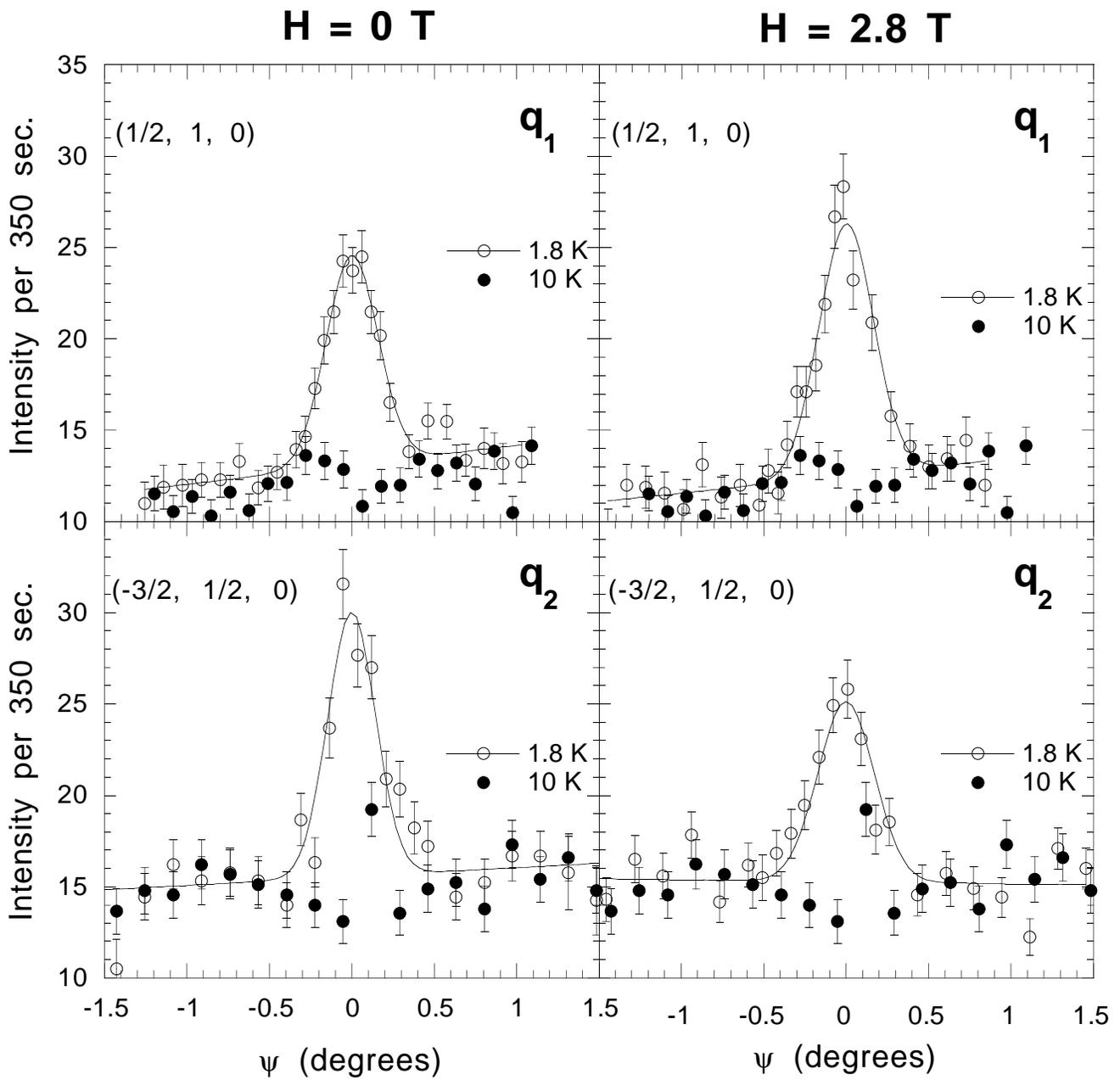
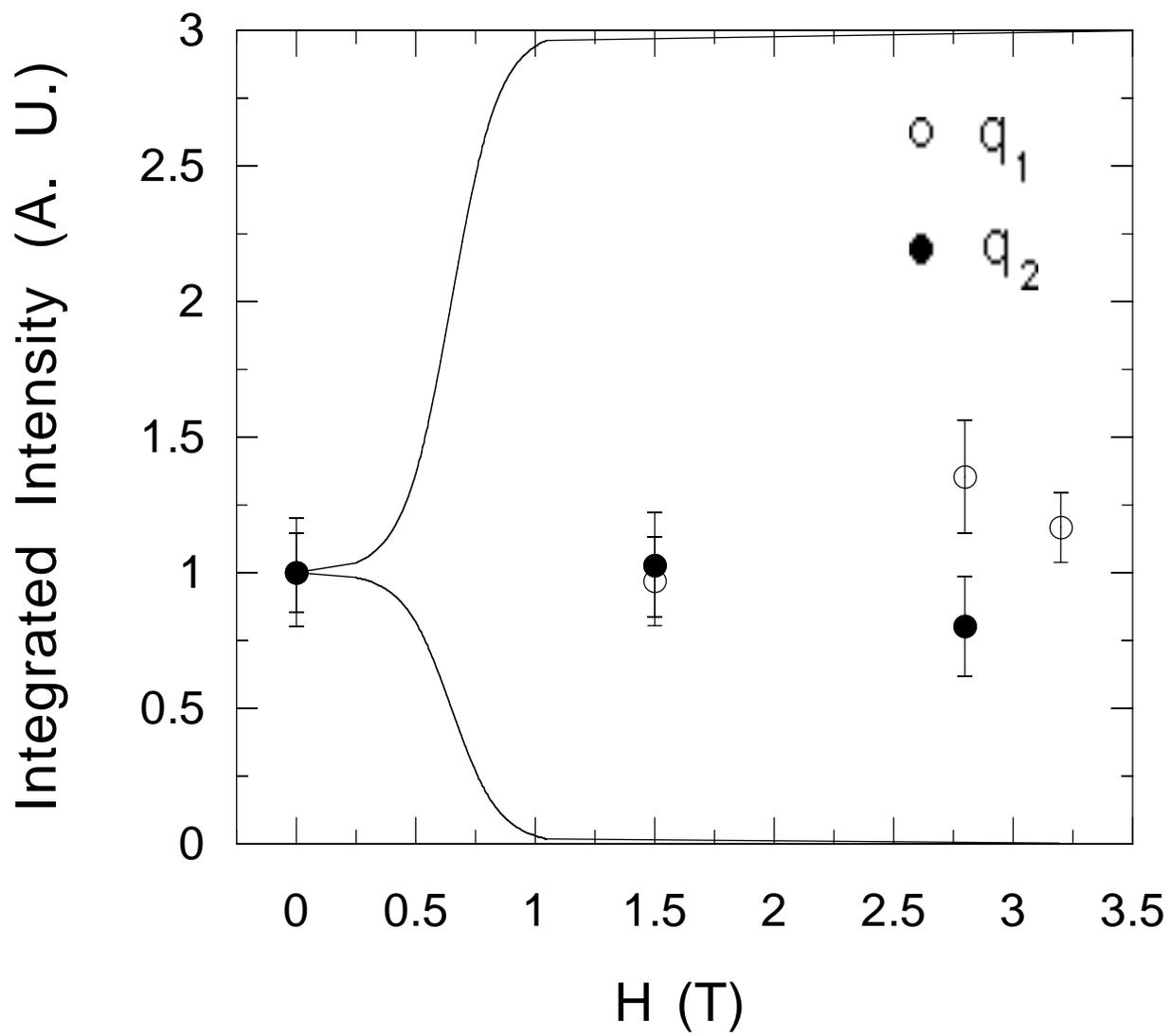
\end{document}